\documentclass[fleqn,usenatbib]{mnras}

\usepackage{times}
\usepackage[T1]{fontenc}
\usepackage{graphicx}
\usepackage{amsmath}
\usepackage{amssymb}
\usepackage{flushend}

\newcommand{\Msun}{\ensuremath{\,{\rm M}_\odot}}                  
\newcommand{\Rsun}{\ensuremath{\,{\rm R}_\odot}}                  
\newcommand{\Teff}{\ensuremath{T_{\rm eff}}}                      
\newcommand{\kms}{\,km\,s$^{-1}$}                                 

\newcommand{\gaia}{\textit{Gaia}}
\newcommand{\reff}[1]{#1}
\newcommand{\refff}[1]{#1}

\title[Pulsations and inclination change in VV\,Ori]
      {A $\beta$\,Cephei pulsator and a changing orbital inclination in the high-mass eclipsing binary system VV\,Orionis}

\author[Southworth et al.]
       {John Southworth\,$^{1}$, D.\ M.\ Bowman\,$^2$, K.\ Pavlovski$^3$ \\
        $^1$\,Astrophysics Group, Keele University, Staffordshire, ST5 5BG, UK \\
        $^2$\,Institute of Astronomy, KU Leuven, Celestijnenlaan 200D, B-3001 Leuven, Belgium \\
        $^3$\,Department of Physics, Faculty of Science, University of Zagreb, Bijenicka cesta 32, 10000 Zagreb, Croatia \vspace*{-15pt}
        }

\date{Accepted 2020/12/07. Received 2020/11/27; in original form 2020/07/10}
\pubyear{2020}

\begin{document} \label{firstpage} \pagerange{\pageref{firstpage}--\pageref{lastpage}} \maketitle 

\begin{abstract}
We present an analysis of the high-mass eclipsing binary system VV\,Ori based on photometry from the TESS satellite. The primary star (B1\,V, 9.5\Msun) shows $\beta$\,Cephei pulsations and the secondary (B7\,V, 3.8\Msun) is \reff{possibly} a slowly-pulsating B star. We detect 51 significant oscillation frequencies, including two multiplets with separations equal to the orbital frequency, indicating that the pulsations are tidally perturbed. We analyse the TESS light curve and published radial velocities to determine the physical properties of the system. Both stars are only the second of their pulsation type with a precisely-measured mass. The orbital inclination is also currently decreasing, likely due to gravitational interactions with a third body.
\end{abstract}



\begin{figure*} \includegraphics[width=\textwidth]{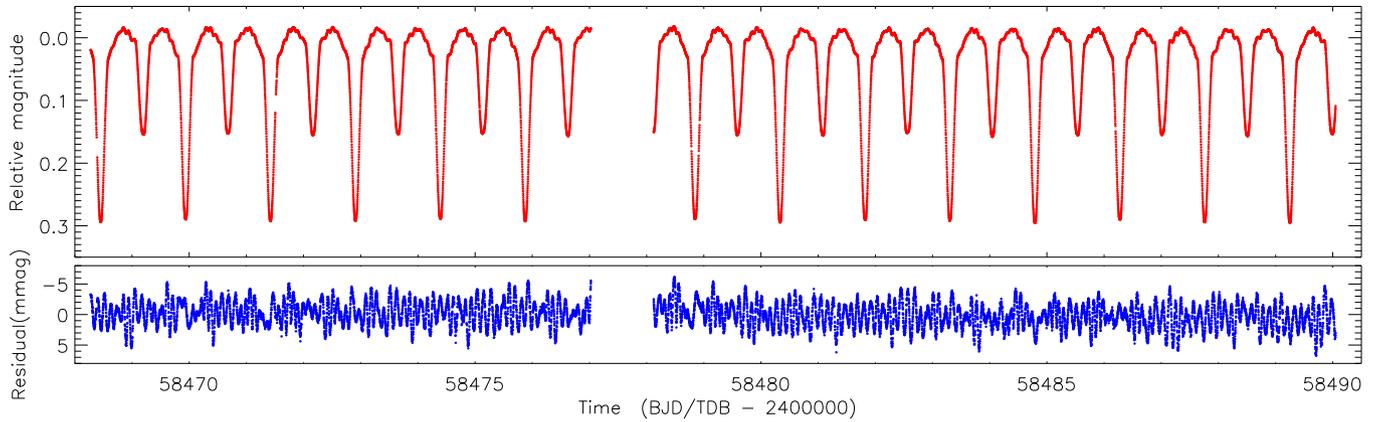}
\caption{\label{fig:tess} Top: TESS simple aperture photometry light curve of VV\,Ori.
Bottom: residuals of the best {\sc jktebop} fit showing the pulsations.} \end{figure*}

\section{Introduction}                                                                                                              \label{sec:intro}

The study of eclipsing binary systems (EBs) is our primary source of empirical measurments of the masses and radii of normal stars \citep{Andersen91aarv,Torres++10aarv}. From light and radial velocity (RV) curves the masses and radii can be determined to high precision and accuracy (reaching 0.2\%; see \citealt{Maxted+20mn}), then used to check and calibrate theoretical models \citep[e.g.][]{ClaretTorres18apj,Tkachenko+20aa} or as distance indicators \citep[e.g.][]{Pietrzynski+19nat}. Massive EBs are of particular interest because massive stars ($\ga$8\Msun) have high multiplicity \citep{Sana+12sci} and are important drivers in the chemical and dynamical evolution of galaxies \citep{Langer12araa}. However, a detailed understanding of their interior rotation and angular momentum transport mechanisms remains elusive \citep{Aerts++19araa}.

Another method to constrain the properties of massive stars is via asteroseismology \citep{Aerts++10book}. Each pulsation mode in a star is sensitive to a specific pulsation cavity, so multi-periodic pulsations are excellent tracers of the properties of stellar interiors \citep[e.g.][]{Aerts+03sci,Briquet+07mn,Briquet+13aa}. Two classes of massive pulsators are the $\beta$\,Cephei stars \citep{LeshAizenman78araa} and the slowly-pulsating B-type (SPB) stars \citep{Waelkens91aa}. The pulsations in $\beta$\,Cephei stars are gravity (g) and pressure (p) modes of low radial order, are found in dwarfs and giants of mass 8--15\Msun, and have amplitudes up to a few tenths of a magnitude and pulsation periods of approximately 2 to 6\,hr \citep{StankovHandler05apjs}. \reff{SPB pulsations occur in stars of roughly 3--9\Msun} and are high-order g-modes with observationally-challenging periods of 1 to 4\,d. Space-based observations have recently revealed p- and g-mode pulsations in many massive stars above their canonical mass regimes, and demonstrated the diverse nature of variability for early-type stars \citep{Pedersen+19apj,Burssens+20aa}.

A promising avenue for constraining the interior physics in massive stars is to study pulsating stars in EBs, as this permits the confrontation of theoretical predictions with observed pulsation periods for stars of known mass and radius. Although several $\beta$\,Cephei stars in EBs are known \citep[see][]{Ratajczak++17conf}, so far only one has a precisely-measured mass and radius: V453\,Cyg\,A \citep{Me+20mn}. SPB stars in EBs are even rarer: only one has been characterised in detail (V539\,Ara; \citealt{Clausen96aa}) and no other examples are known. An important characteristic of binary stars is that tidal effects in eccentric systems may drive \citep{Welsh+11apjs,Hambleton+13mn,Fuller17mn} or perturb \citep{Bowman+19apj,Handler+20natas,Kurtz+20mn,Fuller+20mn} pulsations. 



VV\,Ori contains components of spectral types B1\,V and B7\,V in a circular orbit of period 1.485\,d. Its eclipsing nature was discovered by Barr in 1903 \citep{Barr1905} and the early history of its study has been summarised by \citet{Wood46copri} and \citet{Duerbeck75aas}. The two most recent studies of the system, \citet{SarmaRao95japa} and \citet[][hereafter T07]{Terrell++07mn}, were the first to reliably determine its physical properties.


The multiplicity of the VV\,Ori system is not well established. The excess scatter seen in early RVs led to reluctant suggestions of a third body with an orbital period of 120\,d \citep{Daniel16pallo,StruveLuyten49apj}, a finding that has not been substantiated (T07, \citealt{VanhammeWilson07apj}).  \citet{Horch+17aj} resolved a companion at an angular separation of 0.23$^{\prime\prime}$ using speckle interferometry. The magnitude differences between this companion and the binary system are 3.88\,mag at 692\,nm and 3.43\,mag at 880\,nm. At a distance of $377 \pm 31$\,pc \citep{Gaia18aa} this corresponds to a physical separation of 87\,au and thus a minimum orbital period of about 180\,yr. The resolved companion cannot therefore be responsible for the putative 120\,d orbital variations.

\section{Observations}                                                                                                              \label{sec:obs}

VV\,Ori was observed using the NASA TESS satellite \citep{Ricker+15jatis}. TESS is currently observing the majority of the sky, with each hemisphere divided up into 13 sectors based on ecliptic longitude. Observations in each sector last for 27.4\,d, with an interruption for data download near the midpoint. VV\,Ori was observed 
in Sector 6 (2018/12/11 to 2019/01/07) and is scheduled to be re-observed in Sector 32 (2020/11/19 to 2020/12/17). The Sector 6 data were acquired at a cadence of 2\,min, \reff{which} are made available through the MAST portal, \reff{and yield a Nyquist frequency of 359.7~d$^{-1}$}. 
Fig.\,\ref{fig:tess} shows the simple aperture photometry light curve \citep{Jenkins+16spie} of VV\,Ori.



                                                                                                                                       \label{sec:lc}
\section{Light curve analysis}

\refff{The TESS light curve of VV\,Ori was first modelled using the {\sc jktebop} code in order to remove the signals of binarity (see Fig.\,\ref{fig:tess}). We used the orbital ephemeris from this fit to phase-bin the TESS data into 300 datapoints. The binned data were then analysed using the 2004 version of the Wilson-Devinney code, which represents the stars using Roche geometry \citep[hereafter {\sc wd2004;}][]{WilsonDevinney71apj}, driven by the {\sc jktwd} wrapper \citep{Me+11mn}.} We fitted separately for the light produced by the two stars, the ``third light'' from additional star(s) in the system, the potentials of the two eclipsing stars and the orbital inclination. We did not fit for the effective temperatures (\Teff\ values) directly because the TESS passband is not implemented in the WD code. Limb darkening was included using the square-root law, with coefficients fixed to values from the tables of \citet{Vanhamme93aj}. Fitting for the coefficients did not improve the quality of the fit. We searched for the possibility of an eccentric orbit but found no combination of eccentricity or argument of periastron that improved the fit compared to the assumption of a circular orbit. \refff{We adopted a mass ratio of 0.376 from T07.}

\begin{figure} \includegraphics[width=\columnwidth]{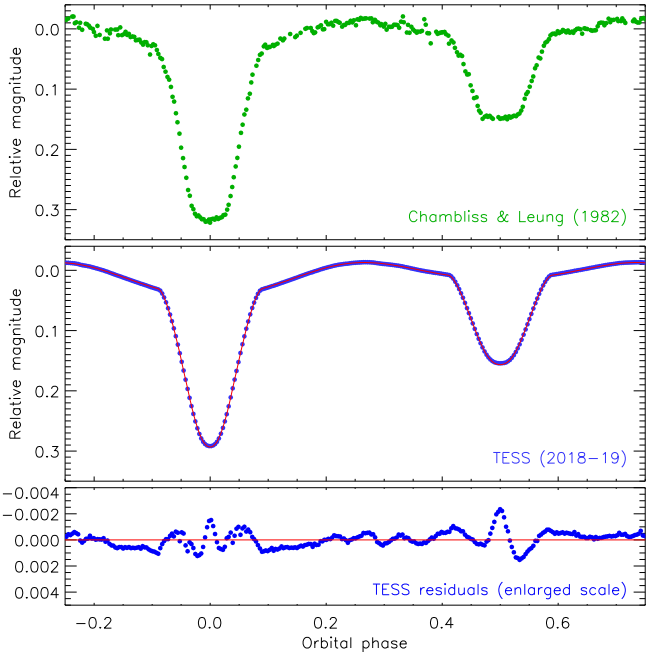}
\caption{\label{fig:phase} Top: light curve from \citet{ChamblissLeung82apjs}
obtained by phase-binning the $BV$ and $vby$ data together. Middle: phase-binned
light curve of VV\,Ori (filled blue circles) with the best fit from the WD code
(red solid line). Bottom: residuals of the fit to the TESS data.} \end{figure}

The adopted model is given in Table\,\ref{tab:absdim} and plotted in Fig.\,\ref{fig:phase}, where the residuals come primarily from incomplete removal of the pulsations. Because the \reff{photon} noise in the photometry is negligible, the uncertainties in the fitted parameters are dominated by choices made in the modelling process. We sought to capture these by quantifying the change in fitted parameter values between the adopted model and a range of alternative models with different values or treatments of albedo, stellar rotation rate, gravity darkening, limb darkening law, limb darkening coefficients, the reflection effect, numerical precision, and mass ratio. All uncertainties in Table\,\ref{tab:absdim} are the quadrature addition of the differences in parameters for each alternative model versus the adopted model, and are much larger than the formal errors computed by {\sc wd2004}.

\section{Physical properties}

Determination of the physical properties of VV\,Ori requires combining the results of the light curve analysis with spectroscopic measurements of the \Teff\ values and velocity amplitudes ($K_1$ and $K_2$) of the two stars. For \Teff\ we adopted the values given by T07 and assigned a conservative uncertainty of $\pm$1000\,K. Improved estimates of the \Teff\ values would be helpful in future.

RVs were measured by T07, who determined the stellar masses but not velocity amplitudes. We therefore fitted the RVs with {\sc jktebop} to determine $K_1 = 126.2 \pm 1.8$\kms\ and $K_2 = 316.4 \pm 6.5$\kms. These uncertainties were calculated using Monte Carlo simulations \citep{Me08mn} and with data errors estimated from the scatter around the best fit. Under the assumption that the orbital inclination decreased by $0.200 \pm 0.013^\circ$\,yr$^{-1}$ in recent years (see below), $K_1$ and $K_2$ should be multiplied by a factor of $(\frac{\sin{i_{2018}}}{\sin{i_{2005}}})^3 = 0.979 \pm 0.002$ to correct them from the mean epoch of the spectra to that of the TESS data. The velocity amplitudes at the epoch of the TESS observations are therefore $K_1 = 123.5 \pm 2.0$\kms\ and $K_2 = 309.6 \pm 7.3$\kms.




The physical properties of the system were then calculated using the {\sc jktabsdim} code \citep{Me++05aa}, which propagates uncertainties using a perturbation analysis. The distance to the system was calculated from the $UBVRI$ magnitudes from \citet{Ducati+01apj}, assuming an uncertainty of 0.02\,mag on each, the $JHK_s$ magnitudes from 2MASS \citep{Skrutskie+06aj}, and the bolometric corrections from \citet{Girardi+02aa}. An interstellar extinction of $E(B-V) = 0.10 \pm 0.05$\,mag was specified to bring the optical and IR distances into agreement. The resulting distance agrees well with the \gaia\ value. 
We find substantially smaller masses for the two stars than T07, and this can be traced to a lower value of $K_2$. 
The uncertainties in both mass and radius are dominated by the contribution from the uncertainty in $K_2$, and also show that the formal errors from the WD code (quoted by T07) are too small. More extensive spectroscopy of VV\,Ori is needed to improve its measured properties.

\begin{table}
\caption{\label{tab:absdim} Brief summary of the parameters for the WD solution of the TESS
light curve of VV\,Ori. Detailed descriptions of the control parameters can be found in
the {\sc wd2004} user guide \citep{WilsonVanhamme04}. Uncertainties are only quoted when
they account for the variation in that parameter over a full set of alternative solutions.}
\setlength{\tabcolsep}{4pt}
\begin{tabular}{@{}lcc@{}} \hline
Parameter                                 & Star A                & Star B                \\
\hline
{\it Control parameters:} \\
{\sc wd2004} operation mode               & \multicolumn{2}{c}{0}                         \\
Treatment of reflection                   & \multicolumn{2}{c}{1}                         \\
Number of reflections                     & \multicolumn{2}{c}{1}                         \\
Limb darkening law                        & \multicolumn{2}{c}{3 (square-root)}           \\
Numerical grid size (normal)              & \multicolumn{2}{c}{60}                        \\
Numerical grid size (coarse)              & \multicolumn{2}{c}{40}                        \\
{\it Fixed parameters:} \\
Orbital period (d)                        & \multicolumn{2}{c}{1.4853784}                 \\
Primary eclipse time (BJD/TDB)            & \multicolumn{2}{c}{2458480.33468}             \\
Mass ratio                                & \multicolumn{2}{c}{0.376}                     \\
Orbital eccentricity                      & \multicolumn{2}{c}{0.0}                       \\
Rotation rates                            & 1.0                   & 1.0                   \\
Bolometric albedos                        & 1.0                   & 1.0                   \\
Gravity darkening                         & 1.0                   & 1.0                   \\
Bolometric linear LD coefficients         & 0.6288                & 0.7121                \\
Linear LD coefficient                     & $-$0.1188             & $-$0.0612             \\
Square-root LD coefficient                & 0.4694                & 0.4073                \\
{\it Fitted parameters:} \\
Orbital inclination (\degr)               & \multicolumn{2}{c}{$78.28 \pm 0.52$}          \\
Third light                               & \multicolumn{2}{c}{$0.005 \pm 0.069$}         \\
Light contributions                       & $11.24 \pm 0.88$      & $1.299 \pm 0.069$     \\
Potentials                                & $3.132 \pm 0.028$     & $3.397 \pm 0.102$     \\
{\it Derived parameters:} \\
Fractional radii                          & $0.3718 \pm 0.0010$   & $0.1817 \pm 0.0034$   \\
Orbital separation (\Rsun)                & \multicolumn{2}{c}{$13.51 \pm 0.05$}          \\
Mass (\Msun)                              & $9.52 \pm 0.56$       & $3.80 \pm 0.16$       \\
Radius (\Rsun)                            & $4.958 \pm 0.088$     & $2.360 \pm 0.061$     \\
Log surface gravity (cgs)                 & $4.026 \pm 0.011$     & $4.272 \pm 0.018$     \\
\Teff\ (K)                                & $26\,200 \pm 1000$    & $16\,100 \pm 1000$    \\
Log luminosity (L$_\odot$)                & $4.02 \pm 0.07$       & $2.53 \pm 0.11$       \\
Absolute bolometric magnitude             & $-5.36 \pm 0.17$      & $-1.65 \pm 0.27$      \\
Distance (pc)                             & \multicolumn{2}{c}{$371 \pm 12$}              \\
\hline
\end{tabular}
\end{table}


\section{Orbital evolution}

Orbital inclination changes have been observed in a small number of EBs; this number is growing due to the multitude of photometric surveys in operation. The list includes V907\,Sco \citep{Lacy++99aj}, SS\,Lac \citep{Torres01aj}, HS\,Hya \citep{ZaschePaschke12aa} and systems discovered \reff{with} the {\it Kepler} satellite \citep{Kirk+16aj} and the OGLE survey \citep{Soszynski+16aca}.

The TESS light curve of VV\,Ori shows deep partial eclipses, but previous photometric studies have revealed the system to display obvious total eclipses \citep[e.g.][]{Eaton75apj,Duerbeck75aas,ChamblissLeung82apjs}. To demonstrate this we show in Fig.\,\ref{fig:phase} a light curve obtained by phase-binning the data obtained in the Johnson $BV$ filters and the Str\"omgren $vby$ filters by \citet{ChamblissLeung82apjs}. The shapes of the two light curves clearly differ. We can rule out a problem with the TESS photometry because a light curve from the MASCARA survey \citep{Burggraaff+18aa} shows similar partial eclipses. Our fit to the TESS data gives an orbital inclination of $i = 78.28 \pm 0.52^\circ$, significantly different to the values of $85.9 \pm 0.2^\circ$ found by T07 from the light curve of \citet{ChamblissLeung82apjs} and $84.5 \pm 0.5^\circ$ found by \citet{Eaton75apj} from his OAO2 data. This suggests that the orbital inclination is changing.

VV\,Ori benefits from a rich observational history stretching back over a century, but unfortunately none of the early light curves are easily accessible. Investigation of possible changes in $i$ are therefore not trivial. We have found two items of evidence that the inclination was lower in the 1940s. Firstly, \citet{Dufay47anap} presents a light curve that shows either partial or marginally total eclipses. His fit to these data, which are rather scattered, gives $i = 77.2^\circ$. Secondly, \citet{HufferKopal50aj} state that the period of totality in primary eclipse lasts for $0.042 \pm 0.003$ phase units. Our by-eye measurement of the duration of totality in the \citet{ChamblissLeung82apjs} light curves is $0.060 \pm 0.005$ in phase units.

Due to the symmetry inherent in spatially-unresolved two-body motion the light curves of EBs have no indication of whether $i$ is above or below 90$^\circ$, so by convention inclinations are quoted as the smaller of $i$ or ($90^\circ-i$). We therefore suggest that VV\,Ori is undergoing orbital precession that manifests as a change in $i$, that it used to show partial eclipses, that the inclination rose to the maximum of $i=90^\circ$ some time in the 1950s or 1960s, and is now decreasing by approximately 0.2$^\circ$\,yr$^{-1}$. Using {\sc wd2004} we find that the lower limit for eclipses to occur in this system is 56.8$^\circ$. When and if VV\,Ori will cease to eclipse in the future will be investigated in detail in a future work. One other relevant constraint is that the binarity of the system was discovered through visual observations of its photometic variability in 1903 \citep{Barr1905}. This almost certainly indicates it was eclipsing at this time because the amplitude of the proximity effects in this system is very low for visual observations: roughly 0.04\,mag at $i = 60^\circ$.


\section{Pulsation analysis}

\begin{figure*} \includegraphics[width=\textwidth]{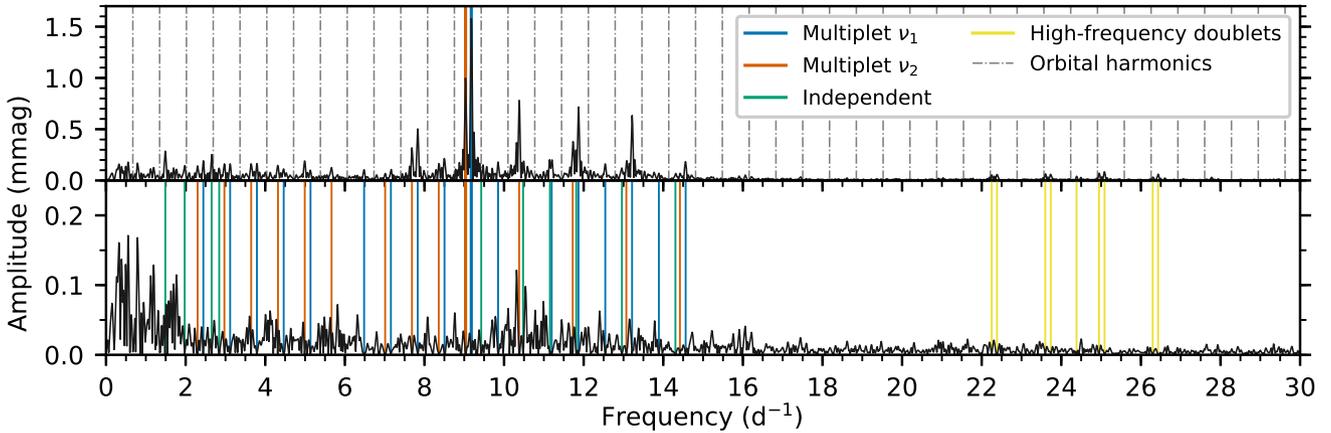}\\[5pt]
\caption{\label{fig:vv:FT} Amplitude spectra of VV\,Ori after removal of the
binary model (top), and after removal of the significant pulsation frequencies
too (bottom). The location and identity of extracted frequencies are shown as
coloured lines, and the dashed grey lines represent \refff{harmonics} of the orbital
frequency.} \end{figure*}

\reff{To} examine the pulsational properties of VV\,Ori, we subtracted a multi-frequency harmonic model \reff{of the orbital frequency,} $\nu_{\rm orb} = 0.673229 \pm 0.000001$~d$^{-1}$, from the light curve and calculated a residual amplitude spectrum using a discrete Fourier transform (\citealt{Kurtz85mn}). We employed iterative pre-whitening to extract all significant frequencies using the standard amplitude signal-to-noise (S/N) criterion of \citet{Breger93apss}, such that significant frequencies have S/N $\geq$ 4. \reff{The S/N of each peak was calculated using its amplitude and the average amplitude within a 1~d$^{-1}$ window at the location of the extracted frequency.} In total we extracted 51 significant frequencies between 1.4 and 27.8\,d$^{-1}$. The frequencies, amplitudes and phases, and their corresponding 1$\sigma$ uncertainties were obtained from a \reff{multi-frequency} non-linear least-squares fit to the light curve \citep{Kurtz+15mn,Bowman17book}, and are provided in Table\,A1. The amplitude spectra \reff{before and after pre-whitening all significant pulsation modes are shown in Fig.\,\ref{fig:vv:FT}.}

VV\,Ori exhibits two multiplets of pulsation modes split by the orbital frequency (centred at $\nu_1 = 9.1766 \pm 0.0001$ and $\nu_2 = 9.0324 \pm 0.0002$~d$^{-1}$), as indicated in Fig.\,\ref{fig:vv:FT}, and several additional pulsation frequencies. \reff{Given the frequencies of the two dominant modes, $\nu_1$ and $\nu_2$, we attribute them to the $\beta$~Cep primary. 
\refff{In slowly-rotating pulsators, symmetric frequency multiplets are plausible \citep[see][]{Bowman20faas}. However, the inferred rotation rate assuming $\nu_1$ and $\nu_2$ as the retrograde and prograde components of a non-radial mode would imply an uncommonly slow rotation rate for such a massive star. On the other hand, such} multiplets are the signature of tidally-perturbed modes \citep{Bowman+19apj,Me+20mn,Jerzykiewicz+20mn,Bowman20faas}, amplitude modulation during the orbit caused by a changing light ratio, and tidally-tilted pulsators \citep{Handler+20natas,Kurtz+20mn,Fuller+20mn}. Our analysis reveals that the pulsation amplitudes are not maximal at a specific orbital phase (see Fig.\,\ref{fig:tess}). Hence, if VV\,Ori is a tidally-tilted pulsator, it is different to those studied by \citet{Handler+20natas} and \citet{Kurtz+20mn}.}

There are several independent g-mode frequencies below 3\,d$^{-1}$, which we suggest arise from the less massive secondary star, \reff{making it a possible SPB star. We also find that the} high-frequency doublets (shown in yellow in Fig.\,\ref{fig:vv:FT}) can be explained by high-order combination frequencies of the orbital frequency and the two dominant modes $\nu_1$ and $\nu_2$ (e.g. $\nu_1 + \nu_2 + n\,\nu_{\rm orb}$ where $n \in \{6,8,10\}$). The remaining significant frequencies are shown in green in Fig.\,\ref{fig:vv:FT}, and we interpret these as independent given that only high-order coincidental combination relations are possible. In Table\,A1 we provide the identifications of the extracted frequencies. Additional `missing' component frequencies of the multiplets can clearly be seen in Fig.\,\ref{fig:vv:FT}, but these fall below our S/N criterion.

\refff{At the suggestion of the referee we investigated whether imperfections in the binary modelling (Sect.\,\ref{sec:lc}) could have affected the pulsation frequencies we have identified in VV\,Ori. To do this we generated a set of slightly different models of the TESS light curve using {\sc jktebop}, by ignoring third light, or fitting for $e\cos\omega$, or fixing the sum of the fractional radii of the stars to a value 4$\sigma$ smaller than the fitted value, or changing the mass ratio and thus the shapes of the stars. The frequency spectra of these datasets show differences at the orbital harmonics, but the pulsation \refff{multiplets} we find are essentially unaffected. This shows that our pulsation analysis is robust against changes in the modelling of binarity\refff{, and that the multiplets are not methodological in origin}.}


\section{Summary and discussion}

VV\,Ori is an early-type binary system containing a 9.5\Msun\ and \ 3.8\Msun\ star on a short-period orbit. It used to have total eclipses but the TESS photometry shows that these eclipses are now partial. We have fitted the TESS light curve and published RVs, and determined the physical properties of the stars. The orbital inclination is significantly lower than found in previous studies, and this change is probably driven by dynamical interactions with a third body. Together with the directly-imaged companion at an angular separation of 0.23$^{\prime\prime}$, this means that the VV\,Ori system is at least quadruple.

The TESS data reveal 51 significant frequencies, including independent g- and p-mode pulsations. We interpret the p-mode pulsations as arising from the primary star, making VV\,Ori\,A the second $\beta$\,Cephei star in an EB with a precisely measured mass (after V453\,Cyg\,A; \citealt{Me+20mn}). The g-mode pulsations \reff{possibly} originate from the secondary star, in which case VV\,Ori\,B is the second known SPB star in an EB (after V539\,Ara\,B; \citealt{Clausen96aa}). \reff{Two frequency multiplets of the dominant p~modes with a spacing of the orbital frequency are also evident.} The equilibrium tide in short-period circular binaries can give rise to a spheroidal bulge and ellipsoidal variability, but is also predicted to perturb self-excited non-radial pulsations and produce \reff{such pulsation frequency} multiplets \citep{ReyniersSmeyers03aa,Balona18mn}. Such multiplets of pulsation modes have recently been detected in a handful of short-period EBs \citep{Bowman+19apj,Handler+20natas,Kurtz+20mn,Me+20mn,Jerzykiewicz+20mn,Fuller+20mn}.

\refff{The multiperiodic frequency spectrum makes VV\,Ori a prime target for future asteroseismic modelling} \citep[see][]{SchmidAerts16aa,Johnston+19mn}. Furthermore, such pulsating stars in EBs open up new avenues to perform tidal asteroseismology and inspect the currently-mysterious interiors of massive stars. In the near future VV\,Ori will be re-observed using TESS, giving a new dataset to further refine the asteroseismic analysis and study the orbital evolution of the system. We have also begun a spectroscopic campaign on this system with the aim of measuring masses and \Teff\ values to higher precision, and detecting \reff{and identifying} line-profile \reff{variability} due to the pulsations in both components.


\section*{Data availability}

All data underlying this article are available in the MAST archive ({https://mast.stsci.edu/portal/Mashup/Clients/Mast/Portal.html}) and from \citet{Terrell++07mn}.

\section*{Acknowledgements}

We thank Gerald Handler and two referees for helpful comments on the manuscript.
The TESS data presented in this paper were obtained from the Mikulski Archive for Space Telescopes (MAST) at the Space Telescope Science Institute (STScI).
STScI is operated by the Association of Universities for Research in Astronomy, Inc. 
Support to MAST for these data is provided by the NASA Office of Space Science. 
Funding for the TESS mission is provided by the NASA Explorer Program.
The research leading to these results has received funding from the Research Foundation Flanders (FWO) by means of a senior postdoctoral fellowship with grant agreement no.\ 1286521N, and from the European Research Council (ERC) under the European Union's Horizon 2020 research and innovation programme (grant agreement no.\ 670519: MAMSIE).


\bibliographystyle{mnras}




\bsp \label{lastpage} \end{document}